\documentclass[letterpaper, 10 pt, conference]{ieeeconf}  %

\IEEEoverridecommandlockouts                              %
\usepackage{graphics} %
\usepackage{epsfig} %
\usepackage{mathptmx} %
\usepackage{times} %
\usepackage{amsmath} %
\usepackage{amssymb}  %

\usepackage[utf8]{inputenc} %
\usepackage[T1]{fontenc}    %

\usepackage{url}            %
\usepackage{booktabs}       %
\usepackage{amsfonts}       %
\usepackage{nicefrac}       %
\usepackage{microtype}      %
\usepackage{xcolor}         %

\usepackage{multirow}

\usepackage{epstopdf}
\usepackage{algorithm,algcompatible}
\usepackage{mathtools}
\usepackage{wrapfig}
\usepackage{soul}
\usepackage{tikz}
\usepackage{xspace}
\usepackage{subcaption}
\usetikzlibrary{fit,positioning,arrows,automata}

\mathtoolsset{showonlyrefs}

\newtheorem{definition}{Definition}
\newtheorem{proposition}{Proposition}
\newtheorem{lemma}{Lemma}

\newcommand{\mE}{{\mathbb E}}

\newcommand{\mR}{{\mathbb R}}

\newcommand{\cE}{{\mathcal E}}

\newcommand{\cL}{{\mathcal L}}
\newcommand{\cN}{{\mathcal N}}
\newcommand{\cO}{{\mathcal O}}

\newcommand{\cX}{{\mathcal X}}
\newcommand{\cY}{{\mathcal Y}}

\def\KL{{\textup{KL}}}
\def\TV{{\textup{TV}}}
\def\LSI{{\textup{LSI}}}

\def\<{{\langle}}
\def\>{{\rangle}}

\def\d{{\text{d}}}

\newcommand{\fan}[1]{{\color{black}{#1}}}
\newcommand{\chen}[1]{{\color{black}{#1}}}

\DeclareMathOperator*{\argmax}{arg\,max}
\DeclareMathOperator*{\argmin}{arg\,min}

\everypar{\looseness=-1}

\title{\LARGE \bf
Nesterov smoothing for sampling without smoothness
}

\author{ Jiaojiao Fan, Bo Yuan, Jiaming Liang and Yongxin Chen%
\thanks{This work was supported by the NSF under grant 1942523 and 2008513.}%
\thanks{J. Fan, B. Yuan and Y. Chen are with the School of Aerospace Engineering,
Georgia Institute of Technology, Atlanta, GA 30332, USA
{\tt\small \{jiaojiaofan,byuan48,yongchen\}@gatech.edu}}%
\thanks{J. Liang is with the Department of Computer Science, Yale University,
        New Haven, CT 06520, USA
        {\tt\small jiaming.liang@yale.edu}}%
}

\begin{document}

\maketitle
\thispagestyle{empty}
\pagestyle{empty}

\begin{abstract}
  We study the problem of sampling from a target distribution in $\mathbb{R}^d$ whose potential is not smooth. Compared with the sampling problem with smooth potentials, this problem is much less well-understood due to the lack of smoothness. In this paper, we propose a novel sampling algorithm for a class of non-smooth potentials by first approximating them by smooth potentials using a technique that is akin to Nesterov smoothing. We then utilize sampling algorithms on the smooth potentials to generate approximate samples from the original non-smooth potentials. 
  We select an appropriate smoothing intensity to ensure that the distance between the smoothed and un-smoothed distributions is minimal, thereby guaranteeing the algorithm's accuracy.
  Hence we obtain non-asymptotic convergence results based on existing analysis of smooth sampling. We verify our convergence result on a synthetic example and apply our method to improve the worst-case performance of Bayesian inference on a real-world example.
\end{abstract}

\section{Introduction}

\chen{Sampling from a target distribution $\pi (x ) \propto \exp(-s(x)) $ known up to a normalization constant is an important problem in many areas such as (particle) filtering and estimation, inverse problems, machine learning, and plays a pivotal role in Bayesian statistics and inference \cite{golightly2006bayesian,stuart2010inverse,murphy2012machine,ghosal2017fundamentals}.
For instance, in nonlinear filtering problem, one can infer from the posterior distribution of the state by sampling from it.
Over the past decades, there has been a vast amount of research carried out on sampling problems from smooth distributions
~\cite{dalalyan2017theoretical,vempala2019rapid}, i.e., the gradient of the potential $s(x)$ is Lipschitz. In contrast, the understanding of non-smooth sampling is still relatively limited.} 

In this work, we consider the task of sampling from a specific class of non-smooth distributions $$\pi(x) \propto \exp(-s(x))$$
whose potentials $s(x)$ permit an explicit max-structure as
\begin{align}\label{eq:potential}
  s(x) = f(x) +  \max_{y \in \cY} \{ \<h(x), y\> -g(y) \}
\end{align}
where $f$ is smooth, $g$ is convex, $\cY$ is convex and bounded, and $h$ is Lipschitz and smooth.
$s(x)$ can be non-smooth.
For instance, when $f(x)=0 $ and $h(x) =Ax$, $s(x)$ becomes a piece-wise affine function and is clearly non-smooth. 
A special instance of particular interest is 
\begin{equation}\label{eq:maxh}
  s(x) = f(x)+ \max_{i} \{h_1(x), h_2(x),\ldots, h_n(x)\}.
\end{equation}
The optimization of minimizing $s(x)$ has been widely used for science and engineering problems to optimize the worst-case performance \cite{ben2009robust,diehl2006approximation}. Its Bayesian counterpart that accounts for uncertainties can be captured by a sampling problem from the potential $s(x)$.

Compared to other recent work ~\cite{liang2021proximal,chewi2021analysis}
for sampling from black box non-smooth distribution, we assume certain structures in the target distribution and hope to leverage these structures to achieve better performance. Indeed, in practice, it is rare to have a black box model as a target distribution; we almost always know some structures of the problem. It could be advantageous to utilize such structures in the algorithm design.
With this in mind, we take one step forward in the direction of sampling beyond black box models and propose a sampling method for non-smooth distribution that utilizes the specific max-structure \eqref{eq:potential}. 

Our strategy is akin to Nesterov smoothing~\cite{nesterov2005smooth} in optimization that constructs a smooth function $s_\beta(x)$ whose difference with $s(x)$ can be controlled through the smoothing intensity $\beta$. In Nesterov smoothing, the idea is to use a fast smooth optimization algorithm to minimize $s_\beta(x)$ for the purpose of optimizing the non-smooth function $s(x)$. In our problem, instead, we sample from $\pi_\beta (x) \propto \exp(-s_\beta(x))$ for the purpose of sampling from $\pi(x) \propto \exp(-s(x))$. Thanks to the smoothness of $s_\beta(x)$, we can take advantage of the many existing sampling algorithms for smooth potentials to sample from $\pi_\beta (x) \propto \exp(-s_\beta(x))$. Our smoothing strategy is compatible with any such algorithms requiring only first-order smoothness. 
The complexity bounds we obtained with Nesterov smoothing have exactly the same dimension dependency as their smooth counterparts. 
In a high level, this work is along the direction of the recent line of research that tries to bridge optimization and sampling~\cite{wibisono2018sampling,barp2022geometric,liang2022proximal}. Our results prove that some smoothing techniques used in optimization are equally effective in sampling.

\paragraph{Related Works}
Over the last few years, several algorithms for sampling from potentials that are not smooth have been developed. A majority of them is devoted to the analysis of original Unadjusted Langevin Monte Carlo (LMC) in the non-smooth setting \cite{erdogdu2021convergence,nguyen2021unadjusted,chewi2021analysis}. These theorectical results are challenging and significantly different from those in the smooth setting. In \cite{bernton2018langevin,durmus2019analysis,salim2020primal}, the non-smooth sampling problem is studied and analyzed from an optimization point of view. Another algorithm that is effective for non-smooth sampling is the proximal sampler developed in the sequence of work \cite{lee2021structured,liang2021proximal,liang2022proximal,chen2022improved}. The methods that are most related to ours are Gaussian smoothing~\cite{chatterji2020langevin}, and Moreau envelope~\cite{durmus2018efficient,lau2022bregman}.
Gaussian smoothing can be applied to any convex potential, however the evaluation of the smoothed potential is difficult. Normally one can only get unbiased estimation of it through Gaussian sampling, inducing additional variance in the algorithm. Moreau envelope can be applied to any weakly-convex potential.
To evaluate the smoothed potential or its gradient, one needs to solve an optimization problem (to get the proximal map), inducing additional complexity, especially when the potential is not convex. We include a more detailed comparison with these two smoothing techniques in the end of Section \ref{sec:error}.

We summarize our contributions below: 
\\
i) Inspired by Nesterov smoothing,
we develop a smoothing technique, different from the existing ones like Gaussian smoothing and Moreau envolepe, for sampling problems for a class of potentials that permit the max-structure \eqref{eq:potential}. Our smoothing technique is compatible with any first-order sampling algorithm for smooth potentials and the convergence is guaranteed via a proper smoothing intensity $\beta$. \\
ii) Combining this smoothing technique with 
existing
sampling algorithms for smooth potentials we obtain non-asymptotic
complexity bounds for strongly-log-concave distributions, log-concave distributions, and distributions satisfying log-Sobolev inequality (LSI). In particular, the complexity has the same dimensional dependency as its smooth counterparts.
 Besides, our analysis covers certain composite potentials with non-smooth (Lipschitz continuous) components and the distribution satisfies LSI. \\
iii) \chen{We show one synthetic example to verify our non-asymptotic analysis and demonstrate that the structure \eqref{eq:potential} can be useful for improving the worst-case performance through a real-world logistic regression example.
}

\section{Background}\label{sec:back}

We assume the norm in the space of $x$ and $y$ can be different and denote them as $\|\cdot \|_\cX$ and $\|\cdot \|_\cY$ respectively.
We use $J_h(x): \mR^d \rightarrow \mR^{n \times d}  $ to represent the  Jacobian matrix of the function $h$ at $x$.
The norm of a matrix $A: \mR^d \rightarrow \mR^{n \times d} $ is defined as $\|A\|_{\cX,\cY} = \max_{x,y} \{
  \< A x, y\>_\cY
  : \|x\|_\cX =1, \|y\|_\cY =1 \}.$
Given a primal space equipped with norm $\|\cdot\|$, we denote the norm in its dual space as $\|\cdot\|^*$.
In this article, we assume the probability measure have density with respect to the Lebesgue measure.

\subsection{Nesterov smooothing}

It is well-known that with subgradient methods, the optimal bound of non-smooth convex minimization complexity is $\cO( 1/\epsilon^2)$ to achieve $\epsilon$ error in the function value~\cite{nemirovskij1983problem,nesterov2003introductory}. However, beyond the black-box oracle model, there are various structures that can be used to improve this bound. Nesterov smoothing~\cite{nesterov2005smooth} is an interesting technique for functions with explicit max structure \eqref{eq:potential}. In doing so, the complexity order can be improved to $\cO(1/\epsilon)$. The underlying principle of Nesterov smoothing is that
 \textit{the dual of a strongly-convex function is smooth and vice versa~\cite{kakade2009duality}.}
Specifically, it adds a $\sigma$-strongly-convex penalty function $\ell(y)$ to the original maximization such that
\begin{align}\label{eq:s_beta}
  s_\beta(x) = f(x) +\max_{y \in \cY} \{\<h(x), y\> -g(y) -
  {\beta} \ell(y) \} ,
\end{align}
where $\beta$ is the smoothing intensity. In the original paper of Nesterov smoothing, they consider the specific task $h(x) =Ax$ and find $s_\beta$ is smooth with constant $L_f + \|A\|^2_{\cX,\cY}/\beta \sigma. $ Then they apply the Nesterov accelerated gradient descent algorithm to minimize $s_\beta$, an optimal scheme for smooth optimization.
Our idea is very similar on the high level: use smoothing for the non-smooth potential first, and then apply the existing sampling algorithm for the smoothed distribution.

\subsection{Smooth sampling}\label{sec:smooth}

In this section, we summarize several popular
sampling algorithms for smooth distributions. In general, there are several dominant popular algorithms: LMC~\cite{parisi1981correlation}, 
Kinetic Langevin Monte Carlo (KLMC)~\cite{cheng2018underdamped} Metropolis-Adjusted Langevin Algorithm (MALA)~\cite{roberts1996exponential}, and Hamiltonian Monte Carlo (HMC)~\cite{neal2011mcmc}.
Suppose we are given the target distribution $\pi \propto \exp(-V(x))$ for some smooth potential $V(x)$.
The LMC algorithm runs the update
\chen{
\begin{align} \label{eq:lmc}
  x_{k+1} = x_{k} - \gamma \nabla V(x_k) + \sqrt{2\gamma} {\epsilon}_k,
\end{align}
where $\epsilon_k$} is the standard Gaussian random variable in $\mR^d$.
The initial point $x_0$ is user-defined and the step size $\gamma>0$ is chosen sufficiently small to ensure limited discretization error.
Indeed, \eqref{eq:lmc} is the Euler discretization of Langevin diffusion dynamics,
\begin{align}\label{eq:langevin}
  \d x_t = - \nabla V(x_t) \d t + \sqrt{2} \d B_t.
\end{align}
This dynamics has an invariant density $\pi$ no matter what $x_0$ is. The underdamped Langevin diffusion adds a Hamiltonian ingredient to the standard Langevin diffusion 
\eqref{eq:langevin}, 
so that an additional random variable $v_t := \d x_t /\d t$ controls the velocity in the dynamics. Similarly, HMC maintains a velocity term, but with a different randomization manner.
MALA is also a variant of LMC. As its name implies, it adds a Metropolis-Hastings rejection step to LMC, in order to eliminate the bias induced by discretization.

In spite of the canonical role of LMC in sampling, it is not the fastest algorithm empirically. Theoretically, in the standard strongly-convex potential case, the complexity of LMC can achieve $\widetilde \cO(d)$~\cite{dalalyan2017theoretical},
whereas KLMC, HMC algorithms can achieve $\widetilde \cO(d^{1/2})$~\cite{dalalyan2020sampling,chen2022optimal}. The complexity $\widetilde \cO(d^{1/2})$ also holds for MALA, however, under a warm start~\cite{wu2021minimax}.
  Decades of extensive research have led to the recent proposal of several innovative sampling algorithms and their variants. Notable examples include the proximal algorithm~\cite{lee2021structured,fan2023improved} based on the restricted Gaussian oracle and the randomized midpoint method for KLMC~\cite{shen2019randomized}. 
  The proximal algorithm  has been proved to have the state-of-art dimension dependency for log-concave distribution~\cite{fan2023improved}.

There exist some accelerating algorithms that can achieve even lower dimension order in complexity~\cite{mangoubi2018dimensionally,sanz2021wasserstein}. However, they require stronger regularity assumptions, e.g. bounded third-order derivative. Although we do not consider these higher-order algorithms in this paper for the sake of generality, one can still apply them in our framework if the higher-order smoothness of $s_\beta(x)$ can be verified in practice.

\section{Nesterov smoothing for sampling}\label{sec:algo}
In this section, we will introduce our Nesterov smoothing sampling algorithm in Section \ref{sec:setup} and bound the error caused by smoothing in Section \ref{sec:error}.
\subsection{Problem setup and algorithm}\label{sec:setup}

Formally, we consider the task to sample from a non-smooth distribution $ \pi(x) \propto \exp(-s(x))$ where
\begin{align}
  s(x) = f(x) +\max_{y \in \cY} \{\<h(x), y\> -g(y) \}.
\end{align}
We will assume $f(x):\cX \rightarrow \mR$ is $L_f$-smooth; $h(x): \cX \rightarrow \cY $ is $\lambda_h$-Lipschitz and $L_h$-smooth; $g(y): \cY \rightarrow \mR$ is continuous and convex through out the paper.
A function $f(x)
$ is $L$-smooth if 
$$\|\nabla f(x_1) -\nabla f(x_2)\|^*_\cX \leq L \|x_1-x_2\|_\cX $$
for $\forall x_1,x_2 \in \mR^d.$
Correspondingly, we say a multivariate function
$h(x)
$ is $L$-smooth if $$\|J_h(x_1) - J_h(x_2) \|_{\cX,\cY} \le L \|x_1 - x_2\|_\cX $$ for~ $\forall x_1,x_2 \in \cX.$
We say a multivariate function $h(x)$ is $\lambda$-Lipschitz if
$$\|h(x_1) -h(x_2) \|^*_\cY \le \lambda \|x_1 -x_2\|_\cX, $$ which is equivalent to $\| J_h(x)  \|_{\cX,\cY} \le \lambda $ ~for~ $\forall x \in \cX$ if $h(x)$ is differentiable.
This boils down to that the
operator norm
of $A$ is bounded when $h(x) =Ax$; this allows us to compare the smoothing effect with the case considered in \cite{nesterov2005smooth}. Without more specifications, 
we assume $\|\cdot\|^*_\cX =\|\cdot\|_\cX =\|\cdot\|_2 $ is Euclidean norm to be consistent with the sampling literature.

It is very likely that $s(x)$ is not differentiable since it is a "piece-wise" function segmented by different maximizer $y$. As such, it is not feasible to directly apply the smooth sampling algorithms.
Inspired by the Nesterov smoothing technique, our strategy is to
 \textit{1) smooth $s(x)$ as $s_\beta(x)$, 2) apply smooth sampling algorithm to $\pi_\beta(x)$.} 

For smoothing, we consider the surrogate function in equation \eqref{eq:s_beta}
where $\ell(y)$ is a $\sigma$-strongly-convex prox-function
$$\ell(y) \ge
  \frac{\sigma}{2}  \| y-y_0\|_\cY^2 .$$
The center of $\cY$ is defined as $y_0 = \argmin_y \{ \ell (y): y \in \cY\} $ and we can assume that $\ell(y_0) =0.$
The compactness of $\cY$ also allows us to
define
\begin{align}\label{eq:bound_l}
  D = \max_y\{\ell(y): y \in \cY \} \text{ , }~~ R = \max_y\{\|y\|_\cY : y \in \cY \}.
\end{align}
From the definition, $D$ is the squared diameter of the set $\cY$, and $R$ is the furthest distance of any point in the set to the origin. By the triangular inequality, it holds that  $$R \le \|y_0\| + \sqrt{\frac{ 2D}{\sigma}}. $$
In this way, $s_\beta(x)$ is a smooth function, which is justified by Lemma \ref{lem:phi_smooth}.
All the proofs can be found in the Appendix.

\vspace{0.2cm}
\begin{lemma}\label{lem:phi_smooth}
  Let $\phi_\beta(x,y): = \<h(x), y\> -g(y) -
    {\beta} \ell(y) $, and $y_\beta(x) := \argmax_{y \in \cY} \phi_\beta(x,y)
  $, then
  \begin{align}\label{eq:grad_s_beta}
    \nabla s_\beta(x) = \nabla f(x) +J_h^\top (x) y_\beta(x).
  \end{align}
  Moreover,
  $s_\beta(x)$ is $L_{s_\beta}$-smooth with
\vspace{-0.3cm}
\begin{align}\label{eq:L_phi}
    L_{s_\beta}= L_f + R L_h + \frac{\lambda_h^2 }{\beta \sigma}.
  \end{align}
\end{lemma}

\vspace{0.2cm}
If $h(x) =Ax$, then $L_h=0$ and $\lambda_h = \|A\|_{\cX,\cY}$, we get $L_{s_\beta} = L_f + \frac{1 }{\beta \sigma} \|A\|^2_{\cX,\cY} $ in this special case. This recovers the smoothness constant in \cite[Theorem 1]{nesterov2005smooth}.
In many cases, $\lambda_h$ is dimensional-free. For example, when $h(x) =Ax $ and $A = c I_m$, then $\lambda_h=c$ regardless of the dimensionality.
 
We will also assume that
we can solve $ \max_y \phi_\beta(x,y)$ for any $x \in \cX$
in a fast dimension-free manner, which has been verified by the complexity analysis~\cite[Section 3]{bubeck2015convex} and especially several examples in \cite[Section 4]{nesterov2005smooth}.
Thus we can query the gradient of $s_\beta(x)$ according to \eqref{eq:grad_s_beta} and
resort to first-order sampling algorithm to sample from $\pi_\beta \propto  \exp(- s_\beta (x))$.  The pseudo-code of our scheme is presented in Algorithm \ref{algo}.

\begin{algorithm}[tb]
  \caption{ Nesterov smoothing sampling framework}
  \label{algo}
  \begin{algorithmic}
    \STATE{{\bfseries Input:}
      A first-order based smooth sampling algorithm,
      the target distribution $\pi \propto \exp(-s(x))$,
      number of iterations $K$,  initial point $x_0 \sim \mu_0$
    }
    \STATE{{\bfseries Initialization:}
      Choose the smoothing intensity $\beta$
    }
    \FOR{ $k=0, \ldots, K-1$}
    \STATE{\it{\# Calculate $\nabla s_\beta(x)  $ according to \eqref{eq:grad_s_beta}} }
    \STATE{Iterate $x_{k+1} $ by the smooth sampling algorithm
    }
    \ENDFOR
    \STATE {{\bfseries Output:}  $x_K $ }
  \end{algorithmic}
\end{algorithm}

\subsection{Bounded distance between $\pi$ and $\pi_\beta$} \label{sec:error}

When $\beta \rightarrow 0,$ it holds that $\pi_\beta \rightarrow \pi $.
In this section, we rigorously bound the distance between $\pi$ and $\pi_\beta$. A key observation is that $s_\beta(x)$ is a uniform smooth approximation of $s(x)$~\cite[Eq. 2.7]{nesterov2005smooth}, i.e., $$ s_\beta(x) \le s(x) \le s_\beta(x)+ \beta D. $$ 
To bound the difference between $\pi$ and $\pi_\beta$, we select two widely-used distributional distances as the error criteria.
The first one is the total variation  $\TV (p,q) = \frac{1}{2} \| p-q\|_1 $, which can be bounded by Kullback–Leibler divergence $\KL (p \|q)$ through Pinsker's inequality
\begin{align}\label{eq:pinsker}
  \TV (p,q) \le \sqrt{\frac{1}{2} \KL(p \| q)}.
\end{align}
The second one is the Wasserstein-2 distance~\cite{villani2021topics}
$$ W_2 (p,q) := \sqrt{\inf_{\nu \in \Pi (p,q)} \int_{\mR^d \times \mR^d } \|a-b\|_2^2 \d\nu(a,b) }, $$
where $\Pi (p,q) $ denotes the set of the all joint distributions of $p$ and $q$.
The $W_2$ distance can also be bounded by $\KL(p \| q)$ through
the Talagrand
inequality~\cite{otto2000generalization}
\begin{align} \label{eq:tala}
  \tag{Talagrand}
  W_2(p,q) \le \sqrt{2 C_\LSI \KL( p\|q)},
\end{align}
if $q$ satisfies the following log-Sobolev inequality with constant $C_\LSI$.

\vspace{0.2cm}
\begin{definition}[Log-Sobolev inequality] \label{def:lsi}
  A probability distribution $q$ satisfies a log-Sobolev inequality with constant $C_\LSI$ if, for all smooth functions $u: \mR^d \rightarrow \mR$, it holds that
  \begin{align}\tag{LSI} \label{eq:lsi}
  \hspace*{-1cm}  \mE_q[u^2 \log{u^2} ] -\mE_q[u^2] \log \mE_q[u^2] \le {2}{C_\LSI} \mE_q [ \|\nabla u\|^2].
  \end{align}
\end{definition}

\vspace{0.2cm}
LSI is a powerful tool for sampling convergence analysis when there is no log-concavity. 
It has nice properties, for instance, it implies the sub-Gaussian concentration property~\cite{ledoux1999concentration}. Moreover, if $s(x)$ is $\alpha$-strongly-convex, then  $\pi(x)$ satisfies the \ref{eq:lsi} with constant $1/\alpha$~\cite{bakry1985diffusions}. With the help of the above inequalities, we show that Proposition \ref{prop:tv}, \ref{prop:w} hold.

\begin{proposition}\label{prop:tv}
  $\TV(\pi , \pi_\beta) \le \frac{\beta D}{2}$.
\end{proposition}
\begin{proposition}\label{prop:w}If $\pi$ satisfies the \ref{eq:lsi} with constant $C_\pi$, then
  $$W_2(\pi , \pi_\beta) \le \sqrt{C_\pi} {\beta D} . $$
\end{proposition}

We next compare the smoothing effect of Nesterov smoothing, Moreau envelope and Gaussian smoothing. 
\chen{Define the non-smooth part in our potential as 
    \[
        \rho(x) = \max_{y \in \cY} \{\<h(x), y\> -g(y) \}
    \]
and denote the Lipschitz constant of $\rho(x)$ as $\lambda_\rho.$}
All three smoothing methods can deal with composite smooth$+$non-smooth potential, but
\emph{Moreau envelope and Gaussian smoothing require the convexity or weakly-convexity of $\rho(x)$,
while Nesterov smoothing does not require so.}
\paragraph{Moreau envelope}
The Moreau-Yosida envelope of a lower semicontinuous (weakly) convex function $\rho(x)$ is defined as 
\begin{align*}
\rho_\beta(x) := \min_{y \in \mR^d} \left\{ \rho(y) + \frac{1}{2\beta} \| x-y\|_2^2 \right\}.
\end{align*}
The smoothed composite potential is then written as $s_\beta(x) = f(x) + \rho_\beta(x).$
If $\rho$ is convex, then $\rho_\beta$ is $L_\rho$-smooth with
$
    L_\rho = \frac{1}{\beta}.
$ 
If $\rho$ is $r$-weakly convex, and $\beta \le 1/(2r)$, the same smoothing effect still holds
\cite{bohm2021variable,hoheiselproximal}. 
According to Proposition 1 in \cite{durmus2018efficient}, 
there is
$
    \TV (\pi, \pi_\beta) \le \frac{1}{2} \beta \lambda_\rho^2,
$
which is implied by $0 \le s(x) -s_\beta(x) \le {\beta  \|\lambda_\rho\|^2}/{2} $.

\paragraph{Gaussian smoothing}
Unlike the proposed method in \cite{chatterji2020langevin}, where they apply Gaussian smoothing to both the smooth part $f(x)$ and the non-smooth part $\rho(x)$ in $s(x)$, our discussion is based on only smoothing the $\rho_\beta(x)$  here. We find 
smoothing both parts would introduce a worse dimension dependency in bounding the difference between $\pi(x)$ and $\pi_\beta(x)$.
 The Gaussian smoothing for $\rho(x)$ is defined as 
 $$\rho_\beta (x) =  \mE_\xi[ \rho ( x + \beta \xi )] ,$$ where $\xi \sim \mathcal{N}(0, I_{d\times d}). $
The smoothing effect of Gaussian smoothing only applies to convex functions. 
 Indeed, according to Lemma 2.2 in~\cite{chatterji2020langevin}, if $\rho(x)$ is convex, then
 $ \rho_\beta (x)$ is $L_{ \rho}$-smooth with
$
   L_{\rho} =  {\lambda_\rho d^{1/2}}/{\beta}.$
Then by Lemma E.2 in \cite{chatterji2020langevin}, 
$
    \TV (\pi, \pi_\beta) \le \frac{1}{2} \beta \lambda_\rho d^{1/2} .
$
Additionally, if $f(x)$ is $\alpha$-strongly convex, by Lemma 3.3 in \cite{chatterji2020langevin},
$
    W_2 (\pi, \pi_\beta) =
    \widetilde{\mathcal{O}}\left(
    \frac{ \beta \lambda_\rho d}{\alpha}
    \right).
$

Now, if we want to achieve $\TV(\pi,\pi_\beta) = \cE/2$, then Moreau envelope results in
$$L_{\rho} = \frac{\lambda_\rho^2}{\cE};$$
Gaussian smoothing needs
$$L_{\rho} = \frac{\lambda_\rho^2 d}{\cE};$$
Nesterov smoothing results in
$$L_{\rho} =  R L_h + \frac{\lambda_h^2 D }{\mathcal{E} \sigma}.$$
Given the same smooth sampling algorithm, the dimension dependency in $L_\rho$ determines the dimension dependency in the final convergence rate.
Normally, $\sigma=1$. Assuming $\lambda_\rho = \lambda_h , ~D =\widetilde{\mathcal{O}}(1) $, then Nesterov smoothing effect is the same order as Moreau envelope, and better than Gaussian smoothing. 
However, Moreau envelope requires more strict assumption on $\rho(x)$, i.e. weakly-convexity; and since the potential is not smooth, solving the proximal operator of Moreau envelope requires additional complexity in each sampling step, even for structured potential in \eqref{eq:potential}.

\section{Non-asymptotic convergence analysis}\label{sec:complexity}
In this section, we present non-asymptotic convergence results in several fundamental scenarios:
strongly log-concavity, log-concavity, and log-Sobolev inequality.
Denote the underlying distribution of $x_{k}$ as $\mu_{k}$. We will give a sufficient number of iterations
for $\mu_k$ to reach a given error tolerance.
In Proposition \ref{thm:str_convex},
each complexity result is accompanied by a 
concrete smoothing algorithm.
As discussed in Section \ref{sec:smooth}, there are multiple algorithm candidates to use for the smooth sampling in Algorithm \ref{algo}. 
We will choose the option with the best complexity to our
knowledge.

To obtain the non-asymptotic convergence results, we notice that the error, for example, $\TV (\mu_K, \pi) $ can be split into two parts $\TV (\mu_K, \pi_\beta) $ and $\TV ( \pi_\beta, \pi)$ by triangular inequality.  The former can be bounded using existing convergence results of smooth sampling, such as \cite{shen2019randomized,fan2023improved}.
Thanks to Proposition \ref{prop:tv}, \ref{prop:w}, the latter is bounded by a carefully chosen $\beta$.
The convergence in terms of $W_2 (\mu_K, \pi) $ follows similarly.

\fan{Under the LSI condition, we will need an additional lemma that guarantees the smoothed distribution $\pi_\beta$ can preserve the LSI property of original $\pi$. 

\vspace{0.2cm}
\begin{lemma}\label{lem:q_lsi}
  If $\pi(x)$ satisfies the \ref{eq:lsi} with constant $C_\pi$, then
  $\pi_\beta$ satisfies the \ref{eq:lsi} with constant
  $C_{\beta} =C_\pi \exp \left(
    4 \beta D
    \right) $.
\end{lemma}

\vspace{0.2cm}

Although under LSI condition, $C_\beta$ has exponential dependence on $D$, we will choose $\beta = \frac{\cE}{D}$ in the algorithm such that $\exp (4 \beta D) = \exp(4\cE) \approx 1 $. In this manner, $C_\beta$ has the same order as $C_\pi$. Precisely, we have the following proposition.}
\vspace{0.3cm}

\begin{proposition}\label{thm:str_convex}
   1) If   $f$ is $\alpha$-strongly-convex, $ \< h(\cdot), y\>$ is convex for any $ y \in \cY$,
 the smooth sampling algorithm in Algorithm \ref{algo} is randomized midpoint KLMC~\cite{shen2019randomized}, and
  $\beta =
    \frac{\sqrt{\alpha} \cE}{2D}
  $,
  then for any $0<\cE< 2\sqrt{\frac{d}{\alpha}}$,
  the iterate $\mu_K$  satisfies $   W_2(\mu_K , \pi) \le \cE$ for
  $K = \widetilde\cO \left(
    \left( \frac{L_{s_\beta}}{\alpha} \right)^{7/6} ( \frac{2}{\cE} \sqrt{\frac{d}{\alpha} } )^{1/3}
    +
    \left( \frac{L_{s_\beta}}{\alpha} \right) ( \frac{2}{\cE} \sqrt{\frac{d}{\alpha} } )^{2/3}
    \right)
  $.
 Furthermore, if $\cE$ is sufficiently small such that $\frac{L_f + R L_h}{\lambda_h^2 D}=
      \cO(1/\cE)$, 
  we obtain
$$K= \widetilde\cO \left(
    \frac{\lambda_h^{2} D d^{1/3} }{\alpha^{11/6} \cE^{5/3}} 
    +    \frac{\lambda_h^{7/3} D^{7/6} d^{1/6} }{\alpha^{23/12} \cE^{3/2}}
    \right) . $$
  
  2) If $f$ is convex, $\< h(\cdot), y\>$ is convex for any $ y \in \cY$, the smooth sampling algorithm is the proximal sampler~\cite{fan2023improved}, and  $\beta =
    \frac{\cE}{D}
  $,  then for any $0< \cE < \cO \left( \frac{\lambda_h^2 D}{L_f + R L_h}  \right) $,
   $\mu_K$  satisfies $ \TV(\mu_K , \pi) \le \cE $ within iterations
   $$K = 
  \cO \left(\frac{\lambda_h^2 D  \sqrt{d} W_2^2(\mu_0, \pi_\beta ) }{ \cE^{3}} \right). 
  $$

  3) If $\pi(x)$ satisfies \ref{eq:lsi} with constant $C_\pi$, the smooth sampling algorithm is the proximal sampler~\cite{fan2023improved}, and $\beta =
    \frac{ \cE}{D}
  $, then for any  $0< \cE < \cO \left( \frac{\lambda_h^2 D}{L_f + R L_h}  \right) 
 $,  $\mu_K$  satisfies $ \TV(\mu_K , \pi) \le \cE $ within   iterations
 $$K = 
  \cO \left(\frac{\lambda_h^2 D C_\pi  \sqrt{d}  }{ \cE } \right) .
  $$
\end{proposition}

\vspace{0.3cm}
The dimension dependence in the above results matches the corresponding convergence result for smooth sampling. 
This is because the smoothing parameter $\beta$ does not depend on the dimension.

\section{Experiments}
In this section, we
consider two examples with the special structure \eqref{eq:maxh}, which can be rewritten as 
\begin{align}
s(x) = f(x) + \max_{y \in \Delta_{n-1}} \{ \< h(x),y \> \},
\end{align}
 where $\Delta_{n-1}$ is the probability simplex on $\mR^n$. To construct the smoothed $s_\beta(x)$, we choose $\| \cdot\|_\cY $ as 1-norm $ \|\cdot\|_1
$, ~$y_0$ is the uniform distribution,
and $\ell(y) = \log (n) + \sum_{j=1}^n y_{j} \log y_{j} $. Thus by \cite[Lemma 3]{nesterov2005smooth}, we have $\sigma=1$ and $D = \log(n)$. We will choose randomized midpoint KLMC as the smooth algorithm.

\subsection{Synthetic example} 
\begin{figure}[h]
  \centering
  \begin{subfigure}{0.4\textwidth}
    \includegraphics[width=0.9\linewidth]{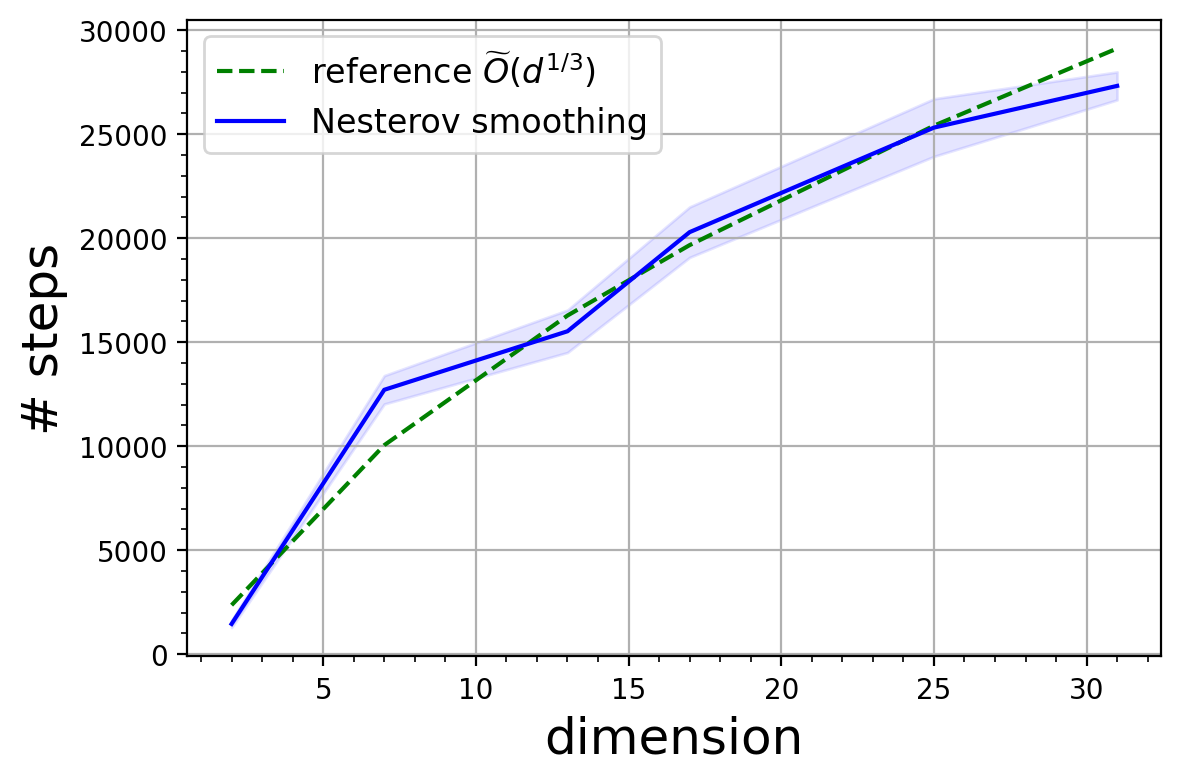}
  \end{subfigure}
  \caption{The number of iterations to reach a fixed error tolerance.
    The reference dashed line is $2700~ d^{\frac{1}{3}}\log(d)$.
  }
  \label{fig:steps}
\end{figure}

We consider $s(x) $ to be the summation of a quadratic term and the maximum of absolute values (also see Section 4.4 of \cite{nesterov2005smooth})
\begin{align}\label{eq:ex}
 \hspace*{-0.5cm} s(x) = \|x\|_\cX^2 +  \max_{1 \le j \le m } \{ |\< a_j , x\> -b_j | \}, \quad a_j \in \mR^d, ~b_j \in \mR.
\end{align}
It is not difficult to find that $\lambda_\rho = \max_{1\le j \le m} \|a_j\|_2$
since it is a piece-wise affine function.
Denote the matrix $A  $ with rows $a_j
$, $b = (b_1, \ldots, b_m)$, and $n=2m.$ 
Then with $\widehat A =
  \begin{pmatrix}
    A \\ -A
  \end{pmatrix}
$
and
$\widehat b =
  \begin{pmatrix}
    b \\ -b
  \end{pmatrix},
$  the form \eqref{eq:ex} is equivalent with
\begin{align}
  s(x) = \|x\|_2^2 +  \max_{y \in \Delta_{n-1} } \{ \< \widehat A x, y\> - \< \widehat b, y\> \}.
\end{align}
Clearly, $s(x)$ is strongly-convex, and we have strongly-convex constant $\alpha=2 $, smoothness $L_f= 2, L_h=0$, and Lipschitz
$$\lambda_h
  = \|\widehat A \|_{\cX,\cY}
  = \max_{x} \{
  \max_{1\le j \le m} |\< a_j, x\>|
  : \|x\|_2 =1 \}
  = \max_{1\le j \le m} \|a_j\|_2 .  
$$
Putting these together in Proposition \ref{thm:str_convex},
we obtain the total complexity 
 $\widetilde \cO \left(
  \frac{
    \lambda_h^{7/3}
    d^{1/3} }{ \cE^{5/3}} \right)$, which is experimentally confirmed in Fig. \ref{fig:steps}.

\subsection{Robust Bayesian logistic regression} 
We present a real-world example of robust Bayesian logistic regression, where
we design a special posterior distribution to make the Bayesian prediction robust to feature perturbation.
Given a dataset $\cL=\{l_1,\ldots,l_S \}$, a model with parameters $x\in \mR^n$ and the prior distribution $p_0(x)$, the nominal task is to sample from the posterior distribution 
$$ \pi_{nom}(x) := p(x|\cL) \propto p_0(x) p(\cL|x). 
$$
Its potential is 
\begin{align}
s_{nom} (x) = - \log p_0(x) -  \log  p(\cL |x) .   
\end{align}
Motivated by \textit{robust approximation problem}~\cite[Sec. 6.4.2]{boyd2004convex} in optimization, we construct another \textit{worse-case} posterior distribution with the potential
\begin{align}
s_{wc}(x) = - \log p_0(x) + \max_i \{ -  \log  p(\cL_i |x) \},  
\end{align}
where $\cL_i$ is the perturbed  version of $\cL$. For example, random noises are added to the features or some features are zeroed. Here we choose to add random Gaussian noise scaled by a noise level to the features,
and a larger $i$ corresponds to a greater noise scale. Our setup can be viewed as a Bayesian counterpart of the optimization problem  $\min ~s_{wc}(x)$, which is a \textit{robust approximation problem}
aiming to minimize the worst-case objective.

We sample two sets of parameters from distributions $\pi_{nom}$ and $\pi_{wc}$, and test their prediction ability on a series of the perturbed test dataset. 
They result in two Bayesian predictions/likelihood
\begin{align}
p_{nom} (l_{test} |\cL )& = \int p( l_{test} | x ) \pi_{nom} (x) \d x , 
\\ 
p_{wc}(l_{test} |\cL )& = \int p( l_{test} | x ) \pi_{wc} (x) \d x 
\end{align}
which are then used to calculate the accuracy (see Sec. \ref{sec:robust_appendix}). In practice, we use posterior samples to estimate the above integrals/expectations.

\begin{figure}[h]
	\centering
	\begin{subfloat}
{\includegraphics[width=0.48\linewidth]{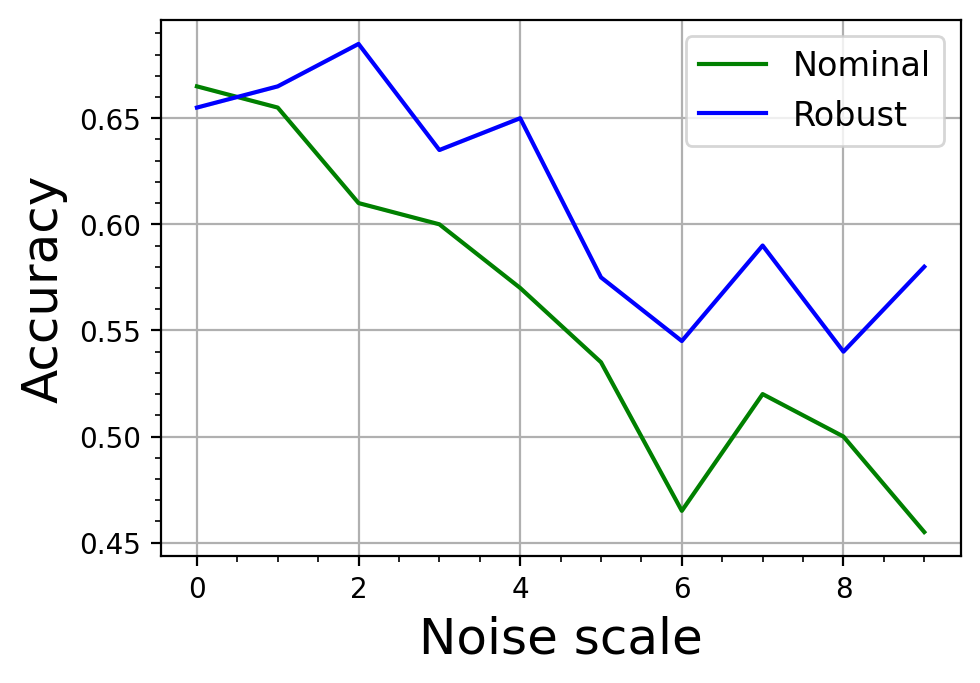}}
	\end{subfloat}
	\begin{subfloat}
		{ \includegraphics[width=0.48\linewidth]{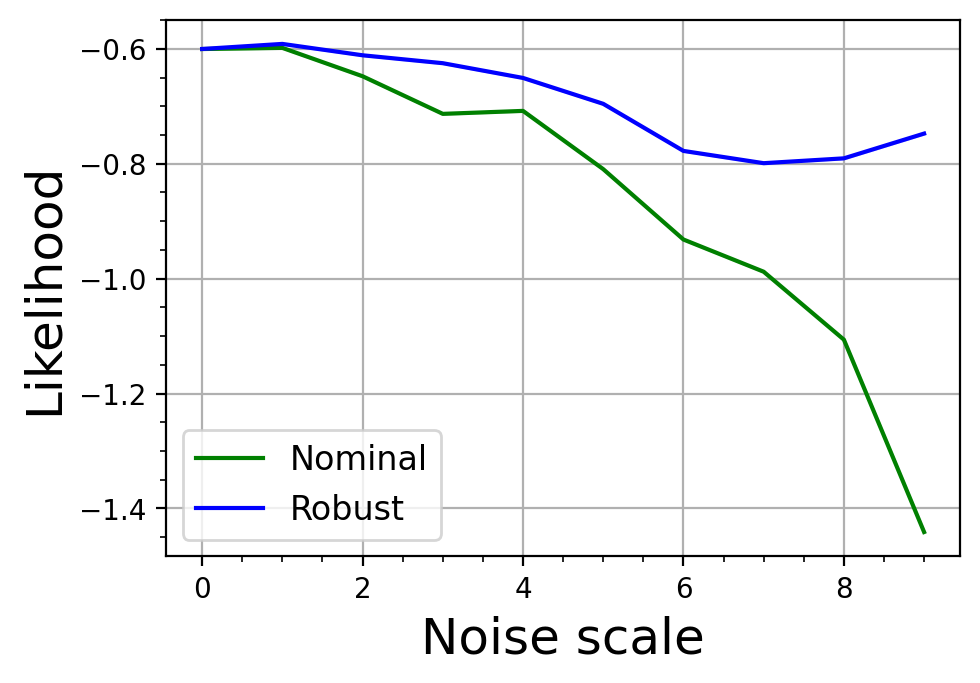}}
	\end{subfloat}
	\caption{Averaged accuracy and the log-likelihood of Bayesian posterior prediction on the test dataset. }
	\label{fig:bayes}
\end{figure}

The dataset $\cL$ we use is "german" dataset from \cite{mika1999fisher}, which includes 1000 data and 20 features. The comparison in Fig. \ref{fig:bayes} illustrates the parameters sampled from $\pi_{wc}$ give a higher accuracy/log-likelihood when there exist data perturbations.

\section{Conclusion}

We consider sampling from a non-smooth distribution $\pi(x) \propto \exp(-s(x))$ that is endowed with the explicit
structure~\eqref{eq:potential}. Inspired by Nesterov smoothing, we first transform the non-smooth potential $s(x)$ to a smooth potential $s_\beta(x)$
and then apply the smooth sampling algorithms directly to $\pi_\beta(x) \propto \exp(- s_\beta(x)) $.
Our smoothing technique is compatible with any first-order sampling algorithm for smooth potentials.
The error introduced by smoothing can be controlled by a careful choice of smoothing intensity $\beta$. 
\fan{We 
study the non-asymptotic convergence rate under several common conditions and present two examples. The logistic regression example reveals our method can be potentially helpful for robust Bayesian inference.}
In this paper, we only consider one class of non-smooth potential with max structure~\eqref{eq:potential} for Nesterov smoothing.
It is interesting to explore more structures that can benefit from smoothing techniques.

{
\bibliographystyle{IEEEtran}
\bibliography{./reference}
}

\appendix

\subsection{Synthetic example}

We normalize $a_j, ~\forall j$ such that $\lambda_h = 4$. We choose the accuracy $\cE = 0.1$, and the dimension of $y $ is $n=10$. In this example,
{with prox-function $\ell(y) = \log (n) + \sum_{j=1}^n y^{(j)} \log y^{(j)} $,} we can solve $y_\beta $ exactly:
\begin{align}
  y_\beta \propto \exp \left(\frac{\widehat A x - \widehat  b}{\beta}  \right).
\end{align}
This aligns with our assumption that we can solve $y_\beta$ in $\cO(1)$. 
Since this example involves a considerable amount of matrix multiplication, we conduct experiments on a GPU card NVIDIA GeForce 2080Ti with 11GB memory.

\subsubsection{Dimension dependency}
Since the total variation error is difficult to calculate in high dimensions, we follow the setup of Section 4.1 in \cite{dwivedi2018log} and
consider using the quantile as the stopping criteria. Specifically, we select $25\%,50\%, 75\%$ quantiles  on the first coordinate direction to compare. Unlike the Gaussian example in \cite{dwivedi2018log}, we do not have access to the ground truth marginal distribution for $x_1$ and can not calculate the real quantile easily. To solve this issue, we follow the experiment setup in \cite{shen2019randomized} and run the sampling algorithm for $K=50000$ steps, which is much larger than the mixing time of this example. Then we take the sample quantile at step $K=50000$ as the ground truth. We use the function \verb+numpy.quantile+ \cite{harris2020array} with 5000 samples to calculate the sample quantile. We use step size 0.1 throughout all dimensions.

We repeat the experiments 5 times with different random seeds and report the error bar in Figure \ref{fig:steps}.

\subsubsection{Trace plot}
\begin{figure}[h]
  \centering
  \begin{subfigure}{1\textwidth}
    \includegraphics[width=0.4\linewidth]{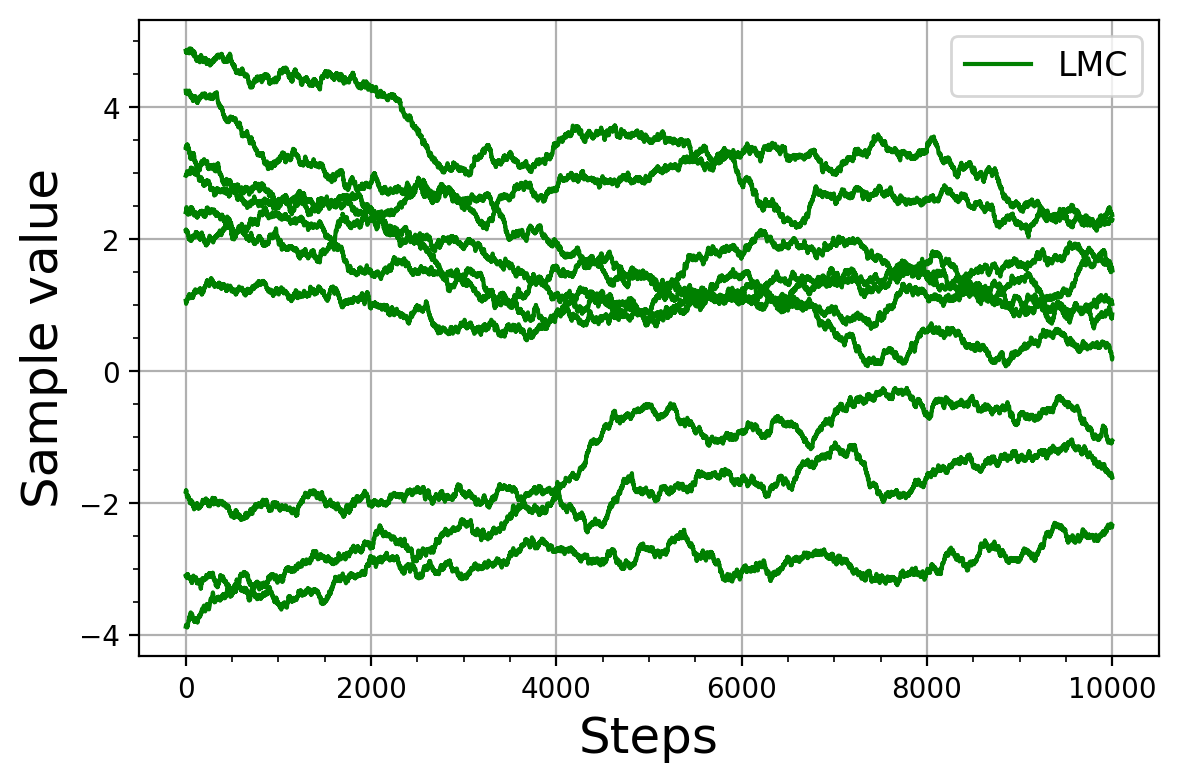}
  
  \includegraphics[width=0.4\linewidth]{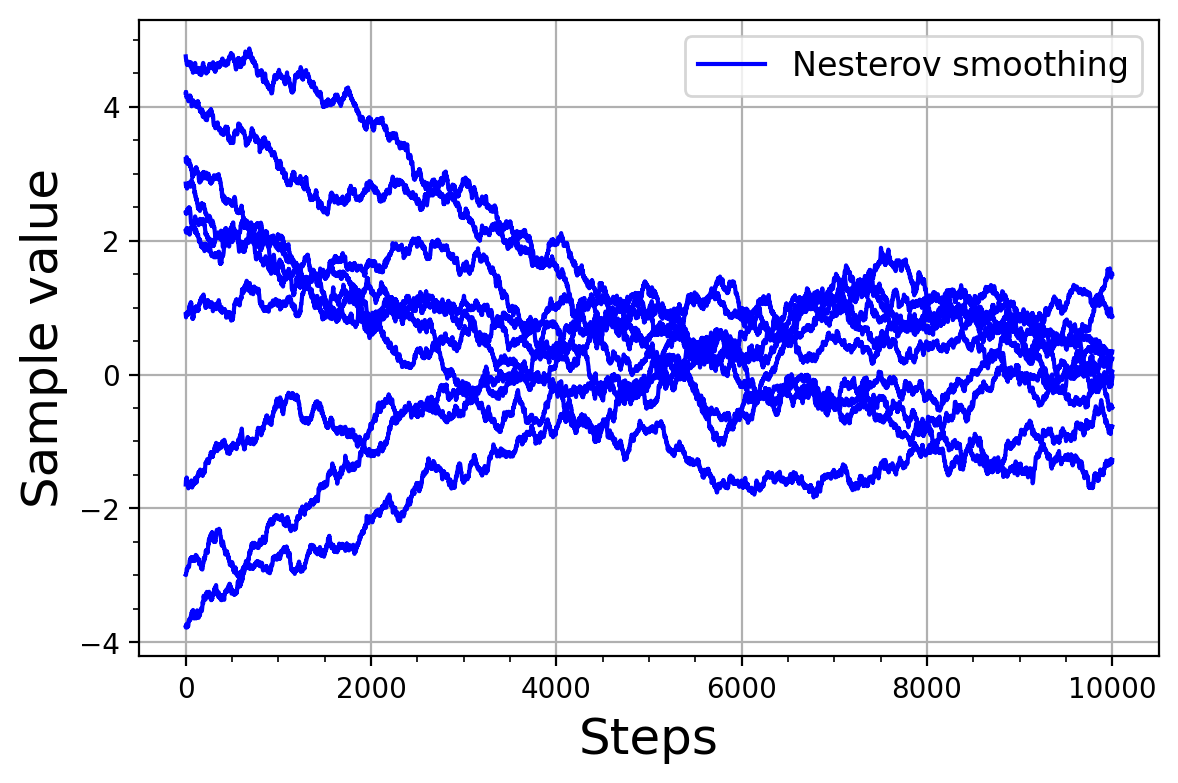}
  \end{subfigure}
  \caption{
    The trajectories of 10 independent runs simulated by LMC and our algorithm at dimension $d=17$.
  }
  \label{fig:trace}
\end{figure}

To investigate the rate of convergence in a single run,
we also give the trace plot of the first coordinate for both our method and Langevin Monte Carlo \eqref{eq:lmc} in Figure \ref{fig:trace}. We conduct the experiments at the dimension $d=17$. The comparison shows that our algorithm converges faster than LMC in high dimension.

\subsection{Robust Bayesian logistic regression}\label{sec:robust_appendix}

The parameter $x $ takes the form of 
{$[\omega,\log \alpha]$}, where $\omega \in \mR^{n-1}$ is the regression weights with the prior $p_0(\omega | \alpha) = \cN(\omega, \alpha^{-1})$. $\alpha$ is a scalar with the prior $p_0(\alpha)= \text{Gamma} (\alpha | 1,0.01)$.

To estimate the log-likelihood and accuracy
of the predictive distribution on $l_{test}$ based
on $\pi_{nom}$ or $\pi_{wc}$, we use straightforward Monte Carlo
estimate on 5000 random $x$ parameter samples.
For accuracy calculation, if $p_{nom} (l_{test} |\cL ) > 0.5$, we consider the prediction to be correct. The same applies when $p_{wc} (l_{test} |\cL ) > 0.5$.

\subsection{Proof of lemmas}

\begin{proof}\textbf{(Lemma \ref{lem:phi_smooth})}
  Define $S(x,y) = f(x) + \<h(x), y\> -g(y) -
    {\beta} \ell(y) $. According to the definition of $\ell(y)$, $S(x,\cdot)$ is $\beta \sigma $-strongly-concave with respect to $\|\cdot\|_\cY$.

  Moreover, since $f$ and $h$ are smooth, we can get
  \begin{align}
    & ~~\| \nabla_x S(x_1,y) - \nabla_x S(x_2,y) \|^*_\cX \\
     & = \|\nabla f(x_1) - \nabla f(x_2) + J_h^\top (x_1) y - J_h^\top (x_2) y \|^*_\cX                \\
     & \le  \|\nabla f(x_1) - \nabla f(x_2) \|^*_\cX + \| J_h^\top (x_1) y - J_h^\top (x_2) y \|^*_\cX \\
     & \le L_f \|x_1 - x_2 \|_\cX +
    \| J_h (x_1)  - J_h (x_2) \|_{\cX,\cY}\| y \|_\cY                                                  \\
     & \le (L_f  +
    L_h R) \|x_1 - x_2 \|_\cX ,\label{eq:Lxx}
  \end{align}
  where we use the definition of matrix norm $\|\cdot \|_{\cX,\cY}$ in the second  last inequality and $\|y\| \le R$ in the last inequality.
  On the other hand, by $h$ is Lipschitz, we have
  \begin{align}
    \| \nabla_x S(x,y_1) - \nabla_x S(x,y_2) \|^*_\cX
     & = \| J_h^\top (x) (y_1 - y_2) \|^*_\cX               \\
     & \le \| J_h^\top (x) \|_{\cX,\cY} \| y_1 - y_2 \|_\cY \\
     & \le \lambda_h \| y_1 - y_2 \|_\cY.\label{eq:Lxy}
  \end{align}
  Similarly, it holds that
  \begin{align}
    \| \nabla_y S(x_1,y) - \nabla_y S(x_2,y) \|
     & = \| h (x_1) -h(x_2) \|
    \le \lambda_h \| x_1-x_2\| . \label{eq:Lyx}
  \end{align}
  Together with \cite[Lemma 1]{sinha2017certifying}, \eqref{eq:Lxx} \eqref{eq:Lxy} \eqref{eq:Lyx} imply that
  $    \nabla s_\beta(x) = \nabla f(x) +J_h^\top (x) y_\beta(x)$ and $\| \nabla s_\beta(x_1) -\nabla s_\beta(x_2) \| \le \left(L_f + R L_h + \frac{\lambda_h^2 }{\beta \sigma} \right)\|x_1 -x_2\|. $
\end{proof}

\begin{proof}\textbf{(Lemma \ref{lem:q_lsi})}
  We can write
  \begin{align}
    \pi_\beta (x) & = \frac{Z_0}{Z_\beta} \exp(-(s_\beta(x) -s(x) ) ) \pi(x) \\
    & = \exp \left(- \left(s_\beta(x) -s(x) + \log \left(\frac{Z_\beta}{Z_0}  \right) \right) \right) \pi(x) ,
  \end{align}
  where $Z_0 = \int_{\mR^d} \exp(-s(x))\d x $ and  $Z_\beta = \int_{\mR^d} \exp(-s_\beta (x))\d x $ are the normalization constants.
  By \eqref{eq:s_diff}, we have  $-{\beta} D \le    s_\beta(x) - s(x) \le 0 ,$  and
  \begin{align}
     & Z_0 \le Z_\beta = \int \exp(-s(x)) \exp(s(x) -s_\beta(x)) \d x\le Z_0 \exp(\beta D) \\
    &  \Rightarrow ~
    0\le \log \left(\frac{Z_\beta}{Z_0} \right) \le \beta D.
  \end{align}
  This implies that $-\beta D \le s_\beta(x) -s(x) + \log \left(\frac{Z_\beta}{Z_0} \right) \le \beta D$.
  Then the result is immediate once notice that
  LSI is stable under bounded perturbation~\cite[Theorem 9.9 (ii)]{villani2021topics}.
\end{proof}

\subsection{Proof of propositions}

\begin{proof} \textbf{(Proposition \ref{prop:tv})}
  According to the definition $\phi_\beta (x,y):= \<h(x), y\> -g(y) - {\beta} \ell(y)$, it's clear that $\phi_\beta(x,y)$ is
  strongly-concave w.r.t. $y$ when $\beta>0$.
  Thus $\phi_\beta(x,y_0 (x))  - \phi_\beta(x,y_\beta (x)) \le 0$.
  This implies
  \begin{align}
    0  \le    s(x) - s_\beta(x)
   & = \phi_\beta(x,y_0 (x)) + {\beta}\ell(y_0(x))  - \phi_\beta(x,y_\beta (x)) \\
    & \le  {\beta}\ell(y_0(x)) \le {\beta} D. \label{eq:s_diff}
  \end{align}
  This further leads to
  \begin{align*}
    \int (s(x) - s_\beta(x))^2 \pi_\beta (x) \d x
    \le \int {\beta^2} D^2 \pi_\beta (x) \d x
    ={\beta^2} D^2.
  \end{align*}
  Then by Pinsker's inequality~\eqref{eq:pinsker} and Lemma 3 in \cite{dalalyan2017theoretical},
  \begin{align*}
    & \TV(\pi , \pi_\beta) \le \sqrt{\frac{1}{2} \KL(\pi_\beta \| \pi )}   \\
    \le & \frac{1}{2} \sqrt{\int (s(x) - s_\beta(x))^2 \pi_\beta (x) dx}
    \le  \frac{\beta D}{2}.
  \end{align*}
\end{proof}

\begin{proof}\textbf{(Proposition \ref{prop:w})}
  According to the definition $\phi_\beta (x,y):= \<h(x), y\> -g(y) - {\beta} \ell(y)$, it's clear that $\phi_\beta(x,y)$ is
  strongly-concave w.r.t. $y$ when $\beta>0$.
  Thus $\phi_\beta(x,y_0 (x))  - \phi_\beta(x,y_\beta (x)) \le 0$.
  This implies
  \begin{align}
    0\le    s(x) - s_\beta(x)
    & = \phi_\beta(x,y_0 (x)) + {\beta}\ell(y_0(x))  - \phi_\beta(x,y_\beta (x)) \\
    & \le  {\beta}\ell(y_0(x)) \le {\beta} D.
  \end{align}
  This further leads to
  \begin{align*}
    \int (s(x) - s_\beta(x))^2 \pi_\beta (x) \d x
    \le \int {\beta^2} D^2 \pi_\beta (x) \d x
    ={\beta^2} D^2.
  \end{align*}
  Then by  Talagrand's inequality and Lemma 3 in \cite{dalalyan2017theoretical},
  \begin{align*}
    & W_2(\pi_\beta , \pi) \le
    \sqrt{{2 C_\pi \KL (\pi_\beta \|\pi)} } \\
  \le & \sqrt{C_\pi  \int (s(x) - s_\beta(x))^2 \pi_\beta (x) dx}
    \le  \sqrt{C_\pi} {\beta D}.
  \end{align*}
\end{proof}

\begin{proof}
  \textbf{(Proposition \ref{thm:str_convex})}
  
  1) Since $ \< h(\cdot), y\>$ is convex for any $ y \in \cY$, it holds that  $ \psi_\beta(x) := \max_{y \in \cY} \{ \<h(x), y\> -g(y)-\beta \ell(y) \} $ is convex for both $\beta>0$ and $\beta=0$. This is because the maximization of the convex function is still convex.

  Remember that we have the relationship: $\alpha$-strongly-convex $s(x) $ implies
  $\pi(x)$ satisfies the \ref{eq:lsi} with constant $1/\alpha$.
  Thus we can bound $W_2(\pi_\beta, \pi ) \le  \beta D /\sqrt{\alpha} $  according to Proposition \ref{prop:w}.
  Plugging in the choice $\beta = \frac{\sqrt{\alpha}\cE}{ 2 D} $ further gives $W_2(\pi_\beta, \pi ) \le \frac{\cE}{2}$.

  On the other hand, with the choice
  \begin{align}\label{eq:K_rm}
    K = \widetilde\cO \left(
    \left( \frac{L_{s_\beta}}{\alpha} \right)^{7/6} \left( \frac{2}{\cE} \sqrt{\frac{d}{\alpha} } \right)^{1/3}
    +
    \left( \frac{L_{s_\beta}}{\alpha} \right) \left( \frac{2}{\cE} \sqrt{\frac{d}{\alpha} } \right)^{2/3}
    \right),
  \end{align}
  we apply the convergence result in Theorem 3 in \cite{shen2019randomized} and have
  $  W_2 (\mu_K , \pi_\beta) \le \frac{\cE}{2}$.
  Putting these two inequalities together gives
  \begin{align}
    W_2(\mu_K , \pi)
    \le  W_2(\mu_K , \pi_\beta)  + W_2(\pi_\beta , \pi)
    \le \frac{\cE}{2} +
    \frac{\cE}{2}
    = \cE.
  \end{align}
  To simplify \eqref{eq:K_rm}, we  plug in $L_{s_\beta} =  L_f + R L_h + \frac{\lambda_h^2 }{\beta \sigma}
    = L_f + R L_h + \frac{2 \lambda_h^2 D }{\sqrt{\alpha} \sigma \cE}  $ and  assume  $\cE$ is sufficiently small such that $\frac{L_f + R L_h}{\lambda_h^2 D}=
    \cO(1/\cE)$. Then $L_{s_\beta} = \cO \left(
    \frac{ \lambda_h^2 D }{\sqrt{\alpha} \sigma \cE}
    \right)$ and
  \begin{align}
       & \widetilde\cO \left(
    \left( \frac{L_{s_\beta}}{\alpha} \right)^{7/6} \left( \frac{2}{\cE} \sqrt{\frac{d}{\alpha} } \right)^{1/3}
    +
    \left( \frac{L_{s_\beta}}{\alpha} \right) \left( \frac{2}{\cE} \sqrt{\frac{d}{\alpha} } \right)^{2/3}
    \right)                    \\
    =~ & \widetilde\cO  \left(
    \frac{\left( \frac{\lambda_h^2 D}{ \sqrt{\alpha  } \sigma \cE } \right)^{7/6}  }{\alpha^{4/3}} \cdot
    \frac{d^{1/6}}{\cE^{1/3}}
    +
    \frac{\frac{\lambda_h^2 D}{ \sqrt{\alpha  } \sigma \cE } }{\alpha \cE^{2/3}} \cdot
    \left(
    \frac{d}{\alpha}
    \right)^{1/3}
    \right)                    \\
    =~ & \widetilde\cO  \left(
    \frac{ \lambda_h^{7/3} D^{7/6} d^{1/6} }{\alpha^{23/12} \cE^{3/2} \sigma^{7/6}}
    +
    \frac{ \lambda_h^{2} D d^{1/3}}{\alpha^{11/6} \cE^{5/3} \sigma }
    \right).
  \end{align}

2)   Since $ \< h(\cdot), y\>$ is convex for any $ y \in \cY$, it holds that  $ \psi_\beta(x) := \max_{y \in \cY} \{ \<h(x), y\> -g(y)-\beta \ell(y) \} $ is also convex for both $\beta>0$ and $\beta=0$. This is because the maximization of the convex function is still convex.

  With the choice of 
     $$K = 
  \cO \left(\frac{
L_{s_\beta}  \sqrt{d} W_2^2(\mu_0, \pi_\beta ) }{ \cE^{2}} \right) , 
  $$ 
  we apply the convergence result in Proposition 11 in \cite{fan2023improved} and have
  $
    \KL (\mu_K \| \pi_\beta) \le {\cE^2}/{2}.
  $
  By Proposition \ref{prop:tv} and the choice $\beta = \frac{\cE}{D} $, there is $ \TV(\pi_\beta . \pi) \le \frac{\cE}{2}. $
  Further, by triangular inequality,
  and Pinsker's inequality,
  \begin{align}
    & \TV(\mu_K , \pi)
    \le  \TV(\mu_K , \pi_\beta)  + \TV(\pi_\beta , \pi) \\
    \overset{\eqref{eq:pinsker} }{\le} & \sqrt{\frac{1}{2} \KL (\mu_K \| \pi_\beta)} + \frac{\cE}{2} \le  \frac{\cE}{2} +
    \frac{\cE}{2}  = \cE.
  \end{align}

  3)   Since  $\pi(x)$ satisfies the \ref{eq:lsi} with constant $C_\pi$, by Lemma \ref{lem:q_lsi}, we have $\pi_\beta$  satisfies the \ref{eq:lsi} with constant $C_\pi \exp(4\beta D)$.

  With the choice
  \begin{align}\label{eq:K_lsi}
    K = \tilde \cO \left( L_{s_\beta} C_\beta \sqrt{d}
    \right),
  \end{align}
  we apply the convergence result in Proposition 13 in \cite{fan2023improved} and have
  $
    \KL (\mu_K \| \pi_\beta) \le {\cE^2}/{2}
  $.
  By Proposition \ref{prop:tv} and the choice $\beta = \frac{\cE}{D} $, there is $ \TV(\pi_\beta . \pi) \le \frac{\cE}{2}. $
  Further, by triangular inequality,
  and Pinsker's inequality,
  \begin{align}
    \TV(\mu_K , \pi)
    \le  \TV(\mu_K , \pi_\beta)  + \TV(\pi_\beta , \pi) \\
    \overset{\eqref{eq:pinsker} }{\le} \sqrt{\frac{1}{2} \KL (\mu_K \| \pi_\beta)} + \frac{\cE}{2} \le  \frac{\cE}{2} +
    \frac{\cE}{2}  = \cE.
  \end{align}
  To simplify \eqref{eq:K_lsi}, we  plug in $L_{s_\beta} =  L_f + R L_h + \frac{\lambda_h^2 }{\beta \sigma}
    = L_f + R L_h + \frac{\lambda_h^2 D }{ \sigma \cE}  $ and  assume  $\cE$ is sufficiently small such that $\frac{L_f + R L_h}{\lambda_h^2 D}=
    \cO(1/\cE)$. Then $L_{s_\beta} = \cO \left(
    \frac{ \lambda_h^2 D }{\sigma \cE}
    \right)$ and $\beta$ is also small enough to do the Taylor approximation $ C_{\beta} =C_\pi \exp \left(
    4 \beta D
    \right)  \approx C_\pi $. Finally, plugging in these pieces results in
  \begin{align}
    K = \tilde \cO \left( L_{s_\beta} C_\beta \sqrt{d} 
    \right)
    =  \widetilde\cO \left(
    \frac{ \lambda_h^2 D C_\pi \sqrt{d} }{\sigma \cE}
    \right).
  \end{align}
  \end{proof}

\end{document}